\documentclass[12pt, a4paper, oneside]{article}
\usepackage[utf8]{inputenc}
\usepackage{authblk}
\usepackage{amsmath}
\usepackage{amssymb}
\usepackage[english]{babel}
\usepackage{epsfig}
\usepackage{float}
\usepackage{cite}
\usepackage[left=3cm,right=1.5cm,top=2cm,bottom=2cm]{geometry}

\begin{document}

\title{ Likelihood Equilibria in the Ising Game \footnote{This work was supported by a grant for research centers in the field of artificial intelligence, provided by the Analytical Center for the Government of the Russian Federation in accordance with the subsidy agreement (agreement identifier 000000D730324P540002) and the agreement with the Moscow Institute of Physics and Technology dated November 1, 2021 No. 70-2021-00138.}}
\author{Andrey Leonidov$^{(a,b)}$}
\affil{{\small 
(a) P.N. Lebedev Physical Institute, Moscow, Russia\\
(b) Moscow Institute of Physics and Technology, Dolgoprudny, Russia\\
 }}

\maketitle

\begin{abstract}
A description of static equilibria in the noisy binary choice (Ising) game on complete and random graphs resulting from maximisation of the likelihood of system configurations is presented.  An equivalence of such likelihood equilibria to the competitive Bayes-Nash quantal response expectation equilibria in the special case of consistent agents expectations is established. It is shown that the same likelihood equilibria are obtained by considering the system's partition function.
\end{abstract}



\section{Introduction}

One of the key challenges in building a quantitative description of multiagent systems is incorporating, to the maximal degree possible, ideas and results from game theory \cite{shoham2008multiagent,lu2020algorithms,vlassis2022concise}. A particularly important role is played by developing various interpretations of game-theoretic equilibria as natural states of mutiagent systems.

A description of static game-theoretic equilibria is in many cases of probabilistic nature. At the fundamental level probabilities provide the most natural interpretation of Nash equilibria in mixed strategies, see e.g. \cite{fudenberg1991game}. Probabilities can also directly appear in the description of a game under consideration. This is in particular the case with games in which agent's utilities include random components. Of direct relevance to the present paper is the corresponding formulation of noisy discrete choice problems \cite{anderson1992discrete} and the particular version of Nash mixed strategies equilibria in such problems, the Quantal Response Equilibria (QRE) \cite{goeree2016quantal,mckelvey1995quantal}. A quantitative description of QRE is based on describing agent's choice as dependent on their expectations with respect to that of neighbouring agents. The QRE can thus be alternatively described as expectation equilibria. Static QRE/expectation equilibria in the noisy binary choice games  were studied in \cite{brock2001discrete,blume2003equilibrium,durlauf2010social,leonidov2020qre, leonidov2024ising}  and, in a more general case of noisy multinomial choice, in \cite{brock2002multinomial,leonidov2024potts}.

A description of a fully heterogeneous game in which one needs to keep track of idiosyncratic characterstics  of every agent is although formally possible but does not allow to construct its interpretable solution. The way out is to include into consideration only a few attributes describing heterogeneity of agent's properites. For example, in graphical games a usual choice is to distinguish agents by a degree of a vertex in which an agent resides, see e.g. \cite{goyal2012connections}. The key quantity reflecting the resulting redundancy is entropy and, indeed it does explicitly appear in describing the dynamical evolution of noisy binary choice (Ising) games towards their equilibria which turn out to be the QRE/expectation ones \cite{blume2003equilibrium,leonidov2024ising} . At the same time available descriptions of static QRE/expectation equilibria are based on describing individual expectations so that the above-mentioned redundancy is reflected only in assuming their appropriate equivalence. In probabilistic terms this means that only expectation values for marginal distributions depending on expectations with respect to neighbour's choices as parameters are considered.

To construct a collective description of game-theoretic equilibria in the presence of random effects one should consider a full multivariate probability distribution of strategies configuration, construct its reduced version appropriate for the problem under consideration and define game-theoretic equilibria in its terms. The most natural candidates for equilibrium configurations are here those maximising the likelihood (i.e. the above-described probability distribution) . In the present paper we continue the analysis of \cite{leonidov2024ising} by explicitly constructing the static likelihood equilibria for the Ising game on complete and random graphs for arbitrary noise and demonstrate their equivalence to entropic equilibria as well as to those constructed through maximising the system's partition function. 

\section{Likelihood Equilibria}

\subsection{Likelihood Equilibria in the Ising Game: General formulae}

An Ising game considered in the present paper is a noisy binary choice game played by $N$ agents placed in the vertices of an undirected graph and equipped with two pure strategies $\{ s_i = \pm 1\}$, $i = 1, \dots ,N$. A general description of such a system is given by the probability distribution (likelihood) ${\cal P} (s_1, \dots , s_N \vert \Theta)$, where $\Theta$ denotes a set of parameters, both idiosyncratic and global, that reflect a particular chosen description of the noisy decision-making process. It is then natural to define a {\it likelihood equilibrium} $(s_1^{\rm eq}, \dots , s_N^{\rm eq})$ of the game under consideration as a configuration having maximal probability of realisation, i.e.
\begin{equation}
(s_1^{\rm eq}, \dots , s_N^{\rm eq}) = \text{arg} \; \underset{s_1, \dots, s_N}{\text{max}} \; {\cal P} (s_1, \dots, s_N \vert \Theta ) 
\end{equation}

An analysis of static equilibria in games on graphs  with noisy discrete response is based on consideration  of expected utilities ${\mathbb E}_{(i)} \left[U_i (s_i) \right] $  reflecting the value of choosing $s_i$ depending in particular on expectations of an agent with respect to choices made by his neighbours. For an Ising game for an agent $i$ the expected utility of choosing the strategy $s_i$ is assumed to have the form \cite{brock2001discrete,blume2003equilibrium,leonidov2024ising}
\begin{equation}\label{uig}
	{\mathbb E}_{(i)} \left[U_i (s_i) \right]   =   \left(H + J \sum_{j \neq i} a_{ij} {\mathbb E}_{(i)} \left[s_j \right]  \right) s_i  + \epsilon_{s_i} 
\end{equation}
where $\{a_{ij}\}$ are matrix elements of the adjacency matrix $a$ with $a_{ij}=1$ corresponding to an edge $j \leftrightarrow i$ and $a_{ij}=0$ otherwise, ${\mathbb E}_{(i)} \left [ s_j \right ]$ stands for the expected value of the choice $s_j$ as perceived by the agent $i$, $J>0$ is a constant characterising mutual influence of agents and $ \epsilon_{s_i} $ are idiosyncratic strategy-dependent random additive contributions to utility with the distribution $f(\epsilon_{s_i} \vert \beta)$ assumed to be the same for all agents where, in turn, the parameter $\beta$ characterises the level of noise. The particular set of parameters $\Theta$ used in this description is thus
\begin{equation}
\Theta = \left(  J, \beta, H, \{  {\mathbb E}_{(i)} \left[s_j \right]  \} \right)
\end{equation}

From equation \eqref{uig} it follows that the probability of choosing the strategy $s_i$ by an agent  $i$ is
\begin{align}\label{psig0}
p (s_i \vert \{ {\mathbb E}_{(i)} \left[s_j \right] \} ) & =  \text{Prob} \left[ 	{\mathbb E}_{(i)} \left[U_i (s_i) \right]  > 	{\mathbb E}_{(i)} \left[U_i (-s_i) \right] \right] \nonumber \\ & =   
F \left( 2 \beta  \left[ H + J \sum_{j \neq i} a_{ij} {\mathbb E}_{(i)} \left[s_j \right]  \right] s_i   \right)
\end{align}
where
\begin{equation}
F(x) = {\rm Prob} \left[ \epsilon_{-s_i} - \epsilon_{s_i}  < x \right] 
\end{equation}
In what follows we will use a log-odds parametrisation of $F(x)$ for which
\begin{equation}\label{parf}
F(x) = \frac{e^{\frac{1}{2}g(x)}}{2 \cosh \left[ \frac{1}{2} g(x) \right]}
\end{equation}
where for symmetric noise distributions $f(\epsilon_{s_i} \vert \beta)$ the function $g(x)$ is an odd function of its argument. In terms of the parametrisation \eqref{parf} equation \eqref{psig0} takes the form
\begin{align}
  & p (s_i \vert \{ {\mathbb E}_{(i)} \left[s_j \right] \} )  =  \label{psig} \\
  & \frac{\exp \left[ s_i \frac{1}{2}  g \left(  2 \beta  \left[ H + J \sum_{j \neq i} a_{ij} {\mathbb E}_{(i)} \left[s_j \right]  \right]   \right) \right]}{2 \cosh  \left[ \frac{1}{2} g \left(  2 \beta  \left[ H + J \sum_{j \neq i} a_{ij} {\mathbb E}_{(i)} \left[s_j \right]  \right]  \right) \right]}, \nonumber
\end{align}
where we have taken into account that $(s_i)^k = s_i$ for $k$ odd and  $(s_i)^k = 1$ for $k$ even. In the special case of the Gumbel noise
\begin{equation}
f(\epsilon_{s_i} \vert \beta) = \beta e^{- \beta \epsilon_{s_i} + e^{- \beta \epsilon_{s_i}}} 
\end{equation}
the function $g(x) = x$ so that 
\begin{equation}\label{psigg}
p (s_i \vert \{ {\mathbb E}_{(i)} \left[s_j \right] \} )  = \frac{\exp \left[    s_i  \left( \beta H + \beta J \sum_{j \neq i} a_{ij} {\mathbb E}_{(i)} \left[s_j \right]  \right)  \right]}{2 \cosh  
\left[  \beta H + \beta J \sum_{j \neq i} a_{ij} {\mathbb E}_{(i)} \left[s_j \right]  \right] }
\end{equation}

For given sets of expectations $\{ {\mathbb E}_{(i)} \left[s_j \right] \}$  for all $i=1, \dots, N$ equation \eqref{psig} does fully describe an equilibrium set of probabilities $\{ p (s_i \vert \{ {\mathbb E}_{(i)} \left[s_j \right] \} ) \}$ defining the corresponding Bayes-Nash competitive equilibrium in mixed strategies. Further simplifications are possible through specifying a structure of the set of expectations. In particular, the quantal response equilibrium \cite{goeree2016quantal,mckelvey1995quantal} is defined by the expectation self-consistency conditions
\begin{equation}\label{qrecond}
{\mathbb E}_{(i)} \left[s_i \right] = {\mathbb E}_{(j)} \left[s_i \right]\;\;\; \forall i,j  
\end{equation}

In the competitive game under consideration  agents form their expectations  independently, so from \eqref{psig} it follows that the likelihood ${\cal P} (s_1, \dots, s_N \vert \{  {\mathbb E}_{(i)} \left[s_j \right]  \})$\footnote{Here and in what follows we omit explicit indication of the parameters $J,\beta,H$ in the formulae for the likelihood.} of a configuration $(s_1, \dots, s_N)$ for a given set of expectations $ \{ {\mathbb E}_{(i)} \left[s_j \right] \}$ equals
\begin{align}\label{likg}
{\cal P} (s_1, \dots, s_N \vert \{  {\mathbb E}_{(i)} \left[s_j \right]  \})   =  \prod_{i=1}^N p (s_i \vert \{ {\mathbb E}_{(i)} \left[s_j \right] \} ) \nonumber \\
 =  \prod_{i=1}^N \frac{\exp \left[ s_i \frac{1}{2} g \left(  2 \beta  \left[ H + J \sum_{j \neq i} a_{ij} {\mathbb E}_{(i)} \left[s_j \right]  \right]   \right) \right]}{2 \cosh  \left[ \frac{1}{2} g \left(  2 \beta  \left[ H + J \sum_{j \neq i} a_{ij} {\mathbb E}_{(i)} \left[s_j \right]  \right]  \right) \right]}
\end{align}
For the Gumbel noise
\begin{align}
& {\cal P} (s_1, \dots, s_N \vert \{  {\mathbb E}_{(i)} \left[s_j \right]  \}) = \\ & \prod_{i=1}^N \frac{\exp \left[    s_i  \left( \beta H + \beta J \sum_{j \neq i} a_{ij} {\mathbb E}_{(i)} \left[s_j \right]  \right)  \right]}{2 \cosh  
\left[  \beta H + \beta J \sum_{j \neq i} a_{ij} {\mathbb E}_{(i)} \left[s_j \right]  \right] } \nonumber
\end{align}
The likelihood equilibrium $(s_1^{\rm eq}, \dots , s_N^{\rm eq})$ of the game under consideration is thus defined by
\begin{equation}
(s_1^{\rm eq}, \dots , s_N^{\rm eq}) = \underset{s_1, \dots, s_N} {\text{arg max}} \; {\cal P} (s_1, \dots, s_N \vert \{  {\mathbb E}_{(i)} \left[s_j \right]  \}) 
\end{equation}

where ${\cal P} (s_1, \dots, s_N \vert \{  {\mathbb E}_{(i)} \left[s_j \right]  \})$ is defined in \eqref{likg}. 

In what follows we consider the properties of likelihood equilibria for the special cases of complete and random graphs, the latter described in the framework of the configuration model.

\subsection{Likelihood Equilibria: Complete Graph}


In the case of complete graph 
\begin{equation}
\sum_{j \neq i} a_{ij} {\mathbb E}_{(i)} \left[s_j \right]    \;\;\; \rightarrow \;\;\; \frac{1}{N} \sum_{j \neq i} {\mathbb E}_{(i)} \left[s_j \right]  
\end{equation}
It is natural to assume that an agent $i$'s expectations with respect to the choices of its $N-1$ neighbours are equivalent, i.e. 
\begin{equation}
{\mathbb E}_{(i)} \left[s_j \right] = m^{(e)}_{(i)}\;\;\; \forall j,
\end{equation}
so that in the limit of large $N$
\begin{equation}
 \frac{1}{N} \sum_{j \neq i} {\mathbb E}_{(i)} \left[s_j \right]  = m^{(e)}_{(i)}
\end{equation}
The likelihood is thus dependent on a set of expectations $\{ m^{(e)}_{(i)}  \}$,
\begin{align}
& {\cal P}(s_1, \dots, s_N \vert \{ m^{(e)}_{(i)}  \})= \label{likcg1} \\ & \prod_{i=1}^N \frac{\exp \left[ s_i \frac{1}{2}  g \left(  2 \beta  \left[ H + J m^{(e)}_{(i)} \right] \right) \right]}{2 \cosh  \left[ \frac{1}{2} g \left(  2 \beta  \left[ H + J m^{(e)}_{(i)}  \right] \right) \right]} \nonumber
\end{align}
In the considered case of the complete graph topology there is nothing idiosyncratic about the agents. It is therefore natural to assume that their expectations  $m^{(e)}_{(i)}$  are (narrowly) centered around some common value $m^{(e)}$\footnote{Let us note that a standard assumption in the literature on graphical games is that expectations of equivalent agents are also equivalent, i.e. that   $m^{(e)}_{(i)} = m^{(e)}$}:
\begin{equation}
m^{(e)}_{(i)} = m^{(e)} + \delta m^{(e)}_{(i)}
\end{equation}
From the structure of \eqref{likcg1} it is clear that in the limit of large $N$ the contribution of possible idiosyncratic expectation shifts $\{ \delta m^{(e)}_{(i)} \}$ is heavily suppressed, so that the expression for the likelihood  \eqref{likcg1} does read
\begin{align}\label{ptotcg1}
& {\cal P}(s_1, \dots, s_N \vert m^{(e)})  =  \prod_{i=1}^N \frac{\exp \left[ s_i \frac{1}{2}  g \left(  2 \beta  \left[ H + J m^{(e)} \right] \right) \right]}{2 \cosh  \left[ \frac{1}{2} g \left(  2 \beta  \left[ H + J m^{(e)}  \right] \right) \right]} \nonumber \\
&= \frac{\exp \left[ N m \frac{1}{2} g(2 \beta [H+J  m^{(e)}]) \right]}{ \left( 2 \cosh \left[ \frac{1}{2} g(2 \beta [H +J  m^{(e)}]) \right] \right)^N}
\end{align}
where we have introduced  an average choice $m$ 
\begin{equation}
m = \frac{1}{N} \sum_{i=1}^N s_i
\end{equation}
As ${\cal P}(s_1, \dots, s_N \vert m^{(e)})$ depends on $(s_1, \dots, s_N)$  only through $m$, the probability distribution \eqref{ptotcg1} can be rewritten as 
\begin{align}\label{ptotcg2}
& {\cal P}(s_1, \dots, s_N \vert m^{(e)})   \to  {\cal P}(m \vert m^{(e)}) \\ & =  {\mathcal N} (m) \frac{\exp \left[ N m \frac{1}{2} g(2 \beta [H+J  m^{(e)}]) \right]}{ \left( 2 \cosh \left[ \frac{1}{2} g(2 \beta [H+J  m^{(e)}]) \right] \right)^N} \nonumber \\
& =  \frac{\exp \left[ N \left( m \frac{1}{2} g(2 \beta [H+J  m^{(e)} ])+ {\mathcal H} (m)  \right) \right]}{ \left( 2 \cosh \left[ \frac{1}{2} g(2 \beta [H+J  m^{(e)}]) \right] \right)^N}, \nonumber
\end{align}
where ${\mathcal N}(m)$ is a number of configurations $(s_1, \dots, s_N)$ such that $(\sum_i s_i)/N = m $ and ${\mathcal H} (m) = \ln {\mathcal N}(m)$  is the Bernoulli entropy
\begin{equation}
{\mathcal H} (m) = - \frac{1-m}{2} \ln \left( \frac{1-m}{2}  \right) - \frac{1+m}{2} \ln \left( \frac{1+m}{2}  \right) 
\end{equation}
We have therefore
\begin{align}\label{ptotcg3}
{\cal P}(m \vert m^{(e)}) & =  \exp \left \{  N \left[ m \frac{1}{2} g(2 \beta [H +J  m^{(e)}])  + {\cal H} (m) \right. \right. \nonumber \\
&-  \left. \left. \ln \left( 2 \cosh \left[ \frac{1}{2} g(2 \beta [H+J  m^{(e)}]) \right] \right) \right] \right \} 
\end{align}

The likelihood equilibrium $m^{\rm eq}$ is defined by
\begin{equation}
m^{\rm eq} = \text{arg} \; \underset{m}{\text{max}} \; {\cal P}(m \vert m^{(e)})
\end{equation}  


The relevant log-likelihood ${\cal V} (m \vert m^{(e)})$ reads 
\begin{align}\label{vcg}
{\cal V} (m \vert m^{(e)}) & = N \left[m \frac{1}{2} g(2 \beta [H+J  m^{(e)}])  + {\cal H} (m) \right] \nonumber \\ & \equiv N {\it v} (m \vert m^{(e)}) 
\end{align}
We have 
\begin{equation}\label{logptotcg}
\frac{d {\it v} (m \vert m^{(e)}) }{d m} = \frac{1}{2} g(2 \beta [H+J  m^{(e)]}) - \text{atanh}(m) 
\end{equation}
The likelihood equilibria $m^{\rm eq}$ do therefore satisfy the equation
\begin{equation}\label{eqeqcg1}
\text{atanh}(m^{\rm eq}) =  \frac{1}{2} g(2 \beta [H+J  m^{(e)}]) 
\end{equation}
or, equivalently,
\begin{equation}\label{eqeqcg2}
 m^{\rm eq} = \tanh \left[ \frac{1}{2} g(2 \beta [H+J  m^{(e)}]) \right] 
\end{equation}
Let us note that 
\begin{equation}
\left. \frac{d^2 {\it v} (m \vert m^{(e)}) }{d m^2} \right \vert_{m=m^{\rm eq}} = - \frac{1}{1-(m^{\rm eq})^2} < 0,
\end{equation}
so that all the found values of  $m^{\rm eq}$ correspond to the maxima of the log-likelihood ${\it v} (m \vert m^{(e)})$.

It is easy to check that the distribution $ {\cal P}(m \vert m^{(e)}) $ defined in \eqref{ptotcg2} is correctly normalised and that
\begin{equation}\label{eqeqcg3}
{\mathbb E}_m [m] = \tanh \left[ \frac{1}{2} g \left(2 \beta [H+J  m^{(e)} ] \right)  \right]
\end{equation}
From from equations (\ref{eqeqcg2},\ref{eqeqcg3}) we see that in the case under consideration
\begin{equation}
m^{\rm eq} = {\mathbb E}_m [m]
\end{equation}

An additional assumption of consistency between the configuration equilibrium and expectations $m^{\rm eq}={\mathbb E}_m [m]=m^{(e)}$ defining expectation/QRE equilibria leads to an equation on $m^{\rm eq}$ of the form
\begin{equation}\label{eqeqcg3}
m^{\rm eq} = \tanh \left[ \frac{1}{2} g(2 \beta [H+J m^{\rm eq}]  ) \right] 
\end{equation} 


The above results can be rederived by analysing, in analogy with the approach developed in statistical physics, the system's partition function. The partition function is computed using unnormalized probabilities of system's configurations. In the case under consideration we get from \eqref{ptotcg2}
\begin{equation}\label{probpfcg}
{\cal P} (m \vert m^{(e)}) \sim  e^{N \left( m\frac{1}{2} g(2 \beta [H+J m^{(e)}]) + {\cal H}(m) \right) } \equiv e^{N v(m \vert m^{(e)})}
\end{equation}
where the function $v(m \vert m^{(e)})$ was defined in \eqref{vcg}
The expression for the partition function ${\cal Z}$ does thus read
\begin{equation}
{\cal Z} = \sum_{m=-1}^1 e^{N v(m \vert m^{(e)})}
\end{equation}
In the partition function-related approach equilibrium states $m^{\rm eq}$ are those giving dominant contribution to ${\cal Z}$. In the limit $N \to \infty$ these are
\begin{equation} 
m^{\rm eq} =  \text{arg} \; \underset{m}{\text{max}} \; v(m \vert m^{(e)}),
\end{equation}
i.e. exactly the same equilibria as described by equation \eqref{eqeqcg2}.

\subsection{Likelihood Equilibria: Random Graph}

Let us now consider likelihood equilibria for the Ising game on undirected random graphs described in the framework of the configuration model, see e.g. \cite{newman2018networks}. In this model a random graph with a given vertex degree distribution $\{ \pi_k \}$ is constructed by randomly creating edges between vertices $i$ and $j$ with degrees $k_i$ and $k_j$ with a probabiliity
\begin{equation}
{\rm Prob} (a_{ij} = 1) = {\mathbb E}[a_{ij}] = \frac{ k_i k_j }{ {\mathbb E}_\pi [k]}
\end{equation}

We will consider the game in the annealed approximation in which we make the following replacement in  \eqref{psig}
\begin{align}\label{repl}
\sum_{j \neq i} a_{ij} {\mathbb E}_{(i)} \left[s_j \right]    \;\;\; & \rightarrow \;\;\;    \sum_{j \neq i} {\mathbb E} [a_{ij}] {\mathbb E}_{(i)} \left[s_j \right]  \nonumber \\ & =
\frac{1}{N} \sum_{j \neq i}  \frac{ k_i k_j }{ {\mathbb E}_\pi [k]} {\mathbb E}_{(i)} \left[s_j \right]  
\end{align}
The only source of heterogeneity in this model is a degree distribution of the vertices. Generically the choice of an agent $i$ is driven by his expectations  with respect to then neighbours choices ${\mathbb E}_{(i)} [s_j] $. In the case under consideration it is assumed, following the standard approach \cite{goyal2012connections},  that an expectation value  ${\mathbb E}_{(i)} [s_j] $ depends only on the degree $k$ of the vertex $j$. In what follows we will use the notation 
\begin{equation}\label{mek}
{\mathbb E}_{(i)} [s_j] = m^{(e)}_k \;\;\;\; \forall i, \;\; k_j = k
\end{equation} 
The probabiility $p_i^{(k)}$ of choosing the strategy $s_i^{(k)}$  by the agent  $i$ located at one of the vertices with the degree $k$ is the same for all such vertices. Using (\ref{repl},\ref{mek}) the argument in \eqref{psig} can be rewritten \cite{leonidov2024ising}  in the following form 
\begin{equation}
H + J \sum_{j \neq i} a_{ij} {\mathbb E}_{(i)} \left[s_j \right]  = H+J k m^{(e)}_w]
\end{equation}
where
\begin{equation}
m^{(e)}_w =    \sum_k \frac{k \pi_k}{ {\mathbb E}_\pi [k]} m^{(e)}_k
\end{equation}
The explicit expression for the probability does therefore read  \cite{leonidov2024ising}
\begin{equation}\label{prg}
p_i ^{(k)}(s_i^{(k)} \vert m^{(e)}_w) = \frac{ e^{\frac{1}{2} s_i^{(k)} g(2 \beta [H+J k m^{(e)}_w])} }{ 2 \cosh \left[ \frac{1}{2} g(2 \beta [H+J k m^{(e)}_w]) \right] } 
\end{equation}

Using the expression \eqref{prg}, we get  the following expression for the distribution function of the system \eqref{likg}:
\begin{align}
& {\cal P}(s_1, \dots, s_N \vert m^{(e)}_w) \\
& = \prod_{k=1}^\infty \frac{\exp \left[ N_k m_k \frac{1}{2} g(2 \beta [H+J  k m^{(e)}_w]) \right]}{ \left( 2 \cosh \left[ \frac{1}{2} g(2 \beta [H+J  k m^{(e)}_w]) \right] \right)^{N_k}}
\nonumber
\end{align}
where $ N_k $ is a number of vertices having degree $k$ and 
\begin{equation}
m_k = \frac{1}{N_k} \sum_{i \in \Xi_k} s_i^{(k)}
\end{equation}
It is convenient to rewrite the distribution in terms of variables $\{ m_k \}$ such that 
\begin{equation}\label{pmk1}
{\cal P}(s_1, \dots, s_N \vert m^{(e)}_w) = \prod_{k=1}^\infty  {\cal P}(m_k \vert m_w^{(e)}) 
\end{equation}
where, in the limit of large $\{ {\cal N}_k \}$
\begin{align}\label{pmk2}
& {\cal P}(m_k \vert m^{(e)})   =  {\cal N} (m_k) \frac{\exp \left[ N_k m_k \frac{1}{2} g(2 \beta [H+J k m_w^{(e)}]) \right]}{ \left( 2 \cosh \left[ \frac{1}{2} g(2 \beta [H+J k m_w^{(e)}]) \right] \right)^{N_k}} \nonumber \\
& =  \frac{\exp \left[ N_k \left( m_k \frac{1}{2} g(2 \beta [H+J  k m_w^{(e)}] )+ {\mathcal H} (m_k)  \right) \right]}{ \left( 2 \cosh \left[ \frac{1}{2} g(2 \beta [H+J  k m_w^{(e)}]) \right] \right)^{N_k}}
\end{align}
Maximisation of  \eqref{pmk2} wiht respect to $m_k$ leads to equations on the equilibrium values of $m_k^{\rm eq}$ describing the corresponding likelihood equilibria. We have
\begin{equation}
\frac{1}{2} g(2 \beta [H+J  k m_w^{(e)}] ) = \text{atanh} (m^{\rm eq}_k) 
\end{equation}
or, equivalently
\begin{equation}\label{mk1}
m^{\rm eq}_k = \tanh \left[ \frac{1}{2} g(2 \beta [H+J  k m_w^{(e)} ]) \right]
\end{equation}
From \eqref{mk1} one gets the following equation for $m^{\rm eq}_w$ 
\begin{align}\label{mw1}
m^{\rm eq}_w & = \sum_k  \frac{k \pi_k}{ {\mathbb E}_\pi [k]} m_k \\ & =  \sum_k  \frac{k \pi_k}{ {\mathbb E}_\pi [k]}  \tanh \left[ \frac{1}{2} g(2 \beta [H+ J  k m_w^{(e)}] ) \right]
\end{align}
In the particular case of QRE $m^{(e)}_k = m^{\rm eq}_k$, so that
\begin{eqnarray}
m^{\rm eq}_k & = & \tanh \left[ \frac{1}{2} g(2 \beta [H+J  k m^{\rm eq}_w ]) \right] \label{mk2} \\
m^{\rm eq}_w & = & \sum_k  \frac{k \pi_k}{ {\mathbb E}_\pi [k]}  \tanh \left[ \frac{1}{2} g(2 \beta [H+J  k m_w]) \right] \label{mw2} \nonumber
\end{eqnarray}

Equations (\ref{mk1},\ref{mw1},\ref{mk2},\ref{mw2}) coincide with those obtained in \cite{leonidov2020qre,leonidov2024ising} where competitive static Bayes-Nash QRE  on random graphs were studied.

\section{Conclusions}

Let us summarise the results obtained in the paper:
\begin{itemize}
\item A description of likelihood equilibria in the Ising game on complete graph is presented and its equivalence to the earlier studied quantal response equilibrium corresponding to a special case of consistent expectations  \cite{leonidov2020qre,leonidov2024ising} is established.
\item A description of likelihood equilibria in the Ising game on random graphs in hte framework of annealed approximation to the configuration model is presented and its equivalence to the earlier studied quantal response equilibrium corresponding to a special case of consistent expectations  \cite{leonidov2020qre,leonidov2024ising} is established.
\item An equivalence of equilibria obtained by considering the system's partition function to the likelihood equilibria for the Ising game on complete graph is established.
\end{itemize}

\end{document}